\documentclass[a4paper,12pt]{article}
\usepackage[english]{babel}

\input epsf.tex

\begin{document}

\centerline {\bf Quantum brachistochrone problem for two spins-$\frac{1}{2}$ with anisotropic Heisenberg interaction}
\medskip
\centerline {A. R. Kuzmak$^1$, V. M. Tkachuk$^2$}
\centerline {\small \it E-Mail: $^1$andrijkuzmak@gmail.com, $^2$voltkachuk@gmail.com}
\medskip
\centerline {\small \it Department for Theoretical Physics, Ivan Franko National University of Lviv,}
\medskip
\centerline {\small \it 12 Drahomanov St., Lviv, UA-79005, Ukraine}

{\small

We study the quantum brachistochrone evolution for a system of two spins-$\frac{1}{2}$ described by an anisotropic Heisenberg Hamiltonian without
$zx$, $zy$ interacting couplings in magnetic field directed along the $z$-axis. This Hamiltonian
realizes quantum evolution in two subspaces spanned by $\vert\uparrow\uparrow\rangle$, $\vert\downarrow\downarrow\rangle$ and
$\vert\uparrow\downarrow\rangle$, $\vert\downarrow\uparrow\rangle$ separately and allows to consider the brachistochrone problem on each subspace separately.
Using the evolution operator for this Hamiltonian we generate quantum gates, namely an entangler gate, SWAP gate, iSWAP gate et al.
\medskip

}

\section{Introduction\label{sec1}}

The quantum brachistochrone problem is formulated similarly to the classical brachistochrone \cite{ref1, ref6, ref2, ref3, ref31, ref4}: What is the optimal
Hamiltonian, under a given set of constraints, such that the evolution from a given initial state $\vert\psi_i\rangle$ to a given final one
$\vert\psi_f\rangle$ is achieved in the shortest time? For the first time this problem was considered by Carlini et al. \cite{ref1}. Using the variational
principle, they presented a general framework for finding the time of optimal evolution and the optimal Hamiltonian for a quantum system with a given set of
initial and final states. The authors solved this problem for some specific examples of constraints. In \cite{ref6} it was shown that analogous results as in \cite{ref1}
can be obtained more directly using symmetry properties of the quantum state space. Their approach was based on the idea considered in \cite{ref7},
where an elementary derivation was provided for the minimum time required to transform an initial quantum state into another orthogonal state. In \cite{ref52}
the quantum brachistochrone problem for mixed states was considered.

Bender et al. explored the brachistochrone problem for a PT-symmetric non-Hermitian Hamiltonian \cite{ref2} and showed that for the non-Hermitian
PT-symmetric Hamiltonians satisfying the same energy constraint the optimal time of evolution between the two fixed states can be made arbitrarily
small. For a more detailed discussion on this subject see \cite{ref3, ref31}. The quantum brachistochrone problem for a spin-1 system in the magnetic
field was solved in \cite{ref8}.

In \cite{ref9, ref10, ref11, ref12} it was established that the entanglement of quantum states is impotant in connection with optimization of quantum
evolution. It was discovered by Giovannetti, Lloyd and Maccone \cite{ref9, ref10} that, in certain cases, entanglement enhances the speed of quantum
evolution of composite systems. Conection between entanglement and time of evolution in the case of two-qubit and n-qubit systems was explored in
\cite{ref13}. The authors of \cite{ref13} showed that, as the number of qubits increases, very little entanglement is needed to reach
the quantum speed limit. Also, it was recently explored that entanglement is an essential resource to achieve the speed limit in the context
of quantum brachistochrone problem \cite{ref14, ref15, ref16}.

In \cite{ref51} it was formulated a
variational principle for finding the optimal time of realization of a target unitary operation, when the available Hamiltonians are subject to certain
constraints dictated by either experimental or theoretical conditions. This method was illustrated for the case of a two-spin system described by an
anisotropic Heisenberg Hamiltonian with $J_{ii}$ interaction between spins ($i=x,y,z$) and magnetic field $h_z^{\alpha}$ ($\alpha=1,2$ for the first and
second spin, respectively) which is directed along the $z$-axis. Also, the authors of \cite{ref51}  generated three examples
of target quantum gates, namely the swap of two qubits ($U_{SWAP}$), the quantum Fourier transformation and the entangler gate ($U_{ENT}$).

Time-optimal generation of quantum gates have already been studied in literature. Recently, speed limits for various
unitary quantum operations in multiqubit systems under typical experimental conditions were obtained \cite{ref23} . Refs.
\cite{ref231, ref232, ref233, ref234, ref235, ref236} contain discusion on the time-optimal generation of unitary operations for a small
number of qubits using Lie group methods, the theory of sub-Riemannian geometry, the Pontryagin maximum principle and assuming that 1-qubit operations can be
performed arbitrarily fast. The time-optimal synthesis of unitary transformation between two
coupled qubits has been discussed in \cite{ref24, ref25, ref26}. The time-optimal control algorithms to synthesize arbitrary unitary transformation for the
coupled fast and slow qubit system were presented in \cite{ref27}. Lower bound on the time required to simulate a two-qubit unitary gate using a
given two-qubit interaction Hamiltonian and local unitaries is provided in \cite{ref28}, while lower bounds on the time complexity of n-qubit gates are given
in \cite{ref29} and upper bounds on the time complexity on certain n-qubit gates are numerically described in \cite{ref30}. A criterion for optimal
quantum computation in terms of a certain geometry in Hamiltonian space was proposed by Nielsen et al. \cite{ref311}. Finally, they showed in \cite{ref32}
that the quantum gate complexity is related to optimal control cost problem.

In \cite{ref17} it was shown that the two-qubit Hamiltonian with the $J_{xx}$ and $J_{yy}$ interaction can generate the iSWAP gate.
Also, the authors showed that applying this gate twice, the CNOT operation can be constructed.
The quantum CNOT gate is the fundamental two-qubit gate for quantum computation \cite{ref18}. This gate plays a central role in networks for quantum error
correction \cite{ref171}. The time-optimal implementation of the CNOT gate on indirectly coupled qubits was studied in \cite{ref22}.

In this paper we consider the quantum brachistochrone problem for the system of two-qubit represented by Heisenberg Hamiltonian with $J_{ii}$ ($i=x,y,z$),
$J_{jk}$ ($j\neq k=x,y$) interaction between spins and magnetic field $h_z^{\alpha}$ ($\alpha=1,2$ for the first and second spin, respectively)
which is directed along the $z$-axis. This Hamiltonian realizes quantum evolution in two subspases spanned by $\vert\uparrow\uparrow\rangle$, $\vert\downarrow\downarrow\rangle$ and
$\vert\uparrow\downarrow\rangle$, $\vert\downarrow\uparrow\rangle$. We solve the quantum brachistochrone problem for each subspase separately and using operator of evolution for
this Hamiltonian we construct entangler, SWAP and iSWAP gates. In comparison with the Hamiltonian from \cite{ref51}, our Hamiltonian contain additional
$J_{xy}$ and $J_{yx}$ couplings which allow to generate the entangler gate along a geodesic.

Thus, this problem might have an important application in quantum computing, quantum teleportation and quantum cryptography,
because, as we are going to show, it allows to reach maximally entangled states during the shortest time.

The paper is organized as follows. In Section \ref{sec2} we introduce and discuss the Hamiltonian of a system of two spins-$\frac{1}{2}$.
In Section \ref{sec3} we solve the quantum brachistochrone problem on the subspace spanned by $\vert\uparrow\uparrow\rangle$, $\vert\uparrow\downarrow\rangle$
(subsection \ref{subsec3_1}) and on the subspace spanned by $\vert\uparrow\downarrow\rangle$, $\vert\downarrow\downarrow\rangle$ (subsection \ref{subsec3_2})
separately and obtain Hamiltonians which provide optimal evolution on each subspaces. We obtain optimal conditions and time required to generate an entangler,
SWAP and iSWAP gates in Section \ref{sec4}. Finally, the summary and discussion are given in Section \ref{sec5}.

\section{Hamiltonian\label{sec2}}

Our aim is to explore quantum brachistochrone evolution for a system of two-qubit. What we consider
is a physical system of a two-qubit represented by two-spin interacting via anisotropic couplings $J_{ii}$ ($i=x,y,z$), $J_{jk}$ ($j\neq k=x,y$) and
magnetic field $h_z^{\alpha}$ ($\alpha=1,2$ for the first and second spin, respectively) which is directed along the $z$-axis.
In other words, we choose the following Hamiltonian:
\begin{eqnarray}
H= \sum_{i, j=x,y,z} J_{ij}\sigma_i^1\sigma_j^2 + \sum_{\alpha=1}^2 h_z^{\alpha}\sigma_z^{\alpha},
\label{form8}
\end{eqnarray}
where $J_{xz}=J_{zx}=J_{yz}=J_{zy}=0$, $\sigma_i^1=\sigma_i\otimes I$, $\sigma_i^2=I\otimes\sigma_i$, $\sigma_i$ are the Pauli matrices.
Note that this Hamiltonian
does not contain items with $\sigma_x^1\sigma_z^2$, $\sigma_z^1\sigma_x^2$ and $\sigma_y^1\sigma_z^2$, $\sigma_z^1\sigma_y^2$,
therefore, it realizes quantum evolution
on two subspaces spanned by $\vert\uparrow\uparrow\rangle$, $\vert\downarrow\downarrow\rangle$ and
$\vert\uparrow\downarrow\rangle$, $\vert\downarrow\uparrow\rangle$ and does not mix these subspaces. This quality allows to rewrite
Hamiltonian (\ref{form8}) as $H=H_I+H_{II}$, where in the basis labeled as $\vert\uparrow\uparrow\rangle$, $\vert\uparrow\downarrow\rangle$,
$\vert\downarrow\uparrow\rangle$, $\vert\downarrow\downarrow\rangle$, the Hamiltonians $H_{I(II)}$ read:
\begin{eqnarray}
H_I&=&\left( \begin{array}{ccccc}
h^++J_{zz} & 0 & 0 & J_{Re}^--iJ_{Im}^+ \\
0 & 0 & 0 & 0 \\
0 & 0 & 0 & 0 \\
J_{Re}^-+iJ_{Im}^+ & 0 & 0 & -h^++J_{zz}
\end{array}\right),\nonumber\\
H_{II}&=&\left( \begin{array}{ccccc}
0 & 0 & 0 & 0 \\
0 & h^--J_{zz}  & J_{Re}^++iJ_{Im}^- & 0 \\
0 & J_{Re}^+-iJ_{Im}^- & -h^--J_{zz} & 0 \\
0 & 0 & 0 & 0
\end{array}\right),
\label{form8_1}
\end{eqnarray}
where we have introduced $h^{\pm}=h_z^1\pm h_z^2$, $J_{Re}^{\pm}=J_{xx}\pm J_{yy}$ and $J_{Im}^{\pm}=J_{xy}\pm J_{yx}$.
$H_I$ and $H_{II}$ commute $\left(\left[H_I,H_{II}\right]=0\right)$. Accordingly, $H_I$ realizes the evolution of a system on the
$\vert\uparrow\uparrow\rangle$, $\vert\downarrow\downarrow\rangle$ subspace and $H_{II}$
realizes the evolution on the $\vert\uparrow\downarrow\rangle$, $\vert\downarrow\uparrow\rangle$ subspace.

Hamiltonian (\ref{form8}) has four eigenvalues: $E_I^+=J_{zz}+\omega_I$, $E_I^-=J_{zz}-\omega_I$, $E_{II}^+=-J_{zz}+\omega_{II}$, $E_{II}^-=-J_{zz}-\omega_{II}$ with the corresponding
eigenvectors:
\begin{eqnarray}
\vert\psi_I^+\rangle &=&\frac{1}{\sqrt{2\omega_I(\omega_I-h^+)}}\left[\left(J_{Re}^--iJ_{Im}^+\right)\vert\uparrow\uparrow\rangle + \left(\omega_{I}-h^+\right)\vert\downarrow\downarrow\rangle\right],\nonumber\\
\vert\psi_I^-\rangle &=&\frac{1}{\sqrt{2\omega_I(\omega_I+h^+)}}\left[\left(J_{Re}^--iJ_{Im}^+\right)\vert\uparrow\uparrow\rangle - \left(\omega_I+h^+\right)\vert\downarrow\downarrow\rangle\right],\nonumber\\
\vert\psi_{II}^+\rangle &=&\frac{1}{\sqrt{2\omega_{II}(\omega_{II}-h^-)}}\left[\left(J_{Re}^++iJ_{Im}^-\right)\vert\uparrow\downarrow\rangle + \left(\omega_{II}-h^-\right)\vert\downarrow\uparrow\rangle\right],\nonumber\\
\vert\psi_{II}^-\rangle &=&\frac{1}{\sqrt{2\omega_{II}(\omega_{II}+h^-)}}\left[\left(J_{Re}^++iJ_{Im}^-\right)\vert\uparrow\downarrow\rangle - \left(\omega_{II}+h^-\right)\vert\downarrow\uparrow\rangle\right],
\label{form8_13}
\end{eqnarray}
where we introduce $\omega_I = \sqrt{{J_{Re}^-}^2+{J_{Im}^+}^2+{h^+}^2}$ and $\omega_{II} = \sqrt{{J_{Re}^+}^2+{J_{Im}^-}^2+{h^-}^2}$.

Hamiltonians $H_I$ and $H_{II}$ have common set of eigenvectors (\ref{form8_13}). $H_I$ has two eigenvalues
$E_I^+$, $E_I^-$ with corresponding eigenvectors $\vert\psi_I^+\rangle$, $\vert\psi_I^-\rangle$ and
one two-fold degenerate eigenvalue $0$ with $\vert\psi_{II}^+\rangle$ and $\vert\psi_{II}^-\rangle$ eigenvectors.
Similar situation is in the case for $H_{II}$. It has two
eigenvalues $E_{II}^+$, $E_{II}^-$ with eigenvectors $\vert\psi_{II}^+\rangle$, $\vert\psi_{II}^-\rangle$, respectively and one
two-fold degenerate eigenvalue $0$ with two eigenvectors $\vert\psi_I^+\rangle$ and $\vert\psi_I^-\rangle$.

By simply reordering the basis states as $\vert\uparrow\uparrow\rangle$, $\vert\downarrow\downarrow\rangle$, $\vert\uparrow\downarrow\rangle$,
$\vert\downarrow\uparrow\rangle$, the Hamiltonian can be rewritten as $H=H_I\oplus H_{II}$, where $H_{I,II}=\left( \begin{array}{ccccc}
h^{\pm}\pm J_{zz} & J_{Re}^{\mp}\mp iJ_{Im}^{\pm} \\
J_{Re}^{\mp}\pm iJ_{Im}^{\pm} & -h^{\pm}\pm J_{zz}
\end{array}\right)$. In this basis, the further notation can be simplified. However, below we will use standard basis labeled as
$\vert\uparrow\uparrow\rangle$, $\vert\uparrow\downarrow\rangle$, $\vert\downarrow\uparrow\rangle$, $\vert\downarrow\downarrow\rangle$ because we
will compare our results with previous papers, in particular, with \cite{ref51} where the standard basis was used.
Also, we will consider quantum gates which are represented in the standard basis.

\section{Quantum brachistochrone \label{sec3}}

The brachistochrone problem is as follows: What is the optimal Hamiltonian, with the finite energy condition, such that the
evolution from a given initial state $\vert\psi_i\rangle$ to a given final one $\vert\psi_f\rangle$ is achieved in the shortest possible time.
We must assume the finite energy condition because the physical systems do not have an unbounded energy resource. For instance, the energy resourse
for a spin-$\frac{1}{2}$ system in the magnetic field is fixed by a value of this field. A simple finite energy condition
is to assume that the difference between the largest and the smallest eigenvalues has a fixed value of $2\omega$:
\begin{eqnarray}
\Delta E=2\omega ,
\label{simpec}
\end{eqnarray}
where $\omega$ is a constant.
We assume the finite energy condition as in \cite{ref51}:
\begin{eqnarray}
{\rm Tr\ H^2} -2\omega^2=0.
\label{energycond}
\end{eqnarray}
For a two-level system with energies $\frac{\Delta E}{2}$ and $-\frac{\Delta E}{2}$ conditions (\ref{simpec}) and (\ref{energycond}) coincide.

Let us provide some introduction on the quntum brachistochrone problem before we consider the quantum brachistochrone for Hamiltonian (\ref{form8}).
In \cite{ref6} a problem on an $n$-dimensional Hilbert space was considered. The authors have shown
that the shortest path joining $\vert\psi_i\rangle$ and $\vert\psi_f\rangle$ should lie on the two-dimensional subspace spaned by
$\vert\psi_i\rangle$ and $\vert\psi_f\rangle$. This subspace is represented
by the Bloch sphere (the Bloch sphere is a sphere of the unit radius which represents the state space of a two-level system). The shortest distance $s_{min}$ between
these states is given by:
\begin{eqnarray}
s_{min}=2\arccos\left(\vert\langle\psi_i\vert\psi_f\rangle\vert\right).
\label{smin}
\end{eqnarray}
The speed $v$ of quantum evolution is given by Anandan-Aharonov relation \cite{ref33}:
\begin{eqnarray}
v=2\sqrt{\langle\psi (t)\vert\Delta H^2\vert\psi (t)\rangle},\nonumber
\end{eqnarray}
where the energy uncertainty $\Delta H$ is bounded by $\omega$. We take $\hbar =1$. The maximum speed of quantum evolution is given by:
\begin{eqnarray}
v_{max}=2\omega.
\label{vmax}
\end{eqnarray}
By using the result in (\ref{smin}) and (\ref{vmax}) we obtain that the minimal time required to realize the unitary transportation
$\vert\psi_i\rangle \rightarrow \vert\psi_f\rangle$ is given by the ratio:
\begin{eqnarray}
t_{min}=\frac{s_{min}}{v_{max}}=\frac{\arccos\left(\vert\langle\psi_i\vert\psi_f\rangle\vert\right)}{\omega}.
\label{tmin}
\end{eqnarray}
The optimal Hamiltonian that generates this unitary transportation is given in \cite{ref6}.

Let us revert to our problem.
The finite energy condition (\ref{energycond}) for Hamiltonian (\ref{form8}) is:
\begin{eqnarray}
{\omega_I}^2+{\omega_{II}}^2+2{J_{zz}}^2=\omega^2.
\label{energyfinite}
\end{eqnarray}

As we mentioned earlier, Hamiltonian (\ref{form8}) realizes evolution in two subspaces separately and does not mix these subspaces due to which we cannot
observe evolution between states from different subspaces. Consequently, we can consider quantum brachistochrone problem on each subspaces separately.

Let us consider the operator of evolution with Hamiltonian (\ref{form8}):
\begin{eqnarray}
U(t)=e^{-iHt}=e^{-iH_It}e^{-iH_{II}t}=U_IU_{II},
\label{form131}
\end{eqnarray}
where we use that $H_I$ and $H_{II}$ commute, and  $U_I(t)=e^{-iH_It}$, $U_{II}(t)=e^{-iH_{II}t}$.
$U_I$ realizes the transformation $\vert\psi_f\rangle = U_I\vert\psi_i\rangle$ on the subspace spanned by $\vert\uparrow\uparrow\rangle$,
$\vert\downarrow\downarrow\rangle$ and acts as a unit operator on another subspace (subspace spanned by $\vert\uparrow\downarrow\rangle$,
$\vert\downarrow\uparrow\rangle$). Contrary to $U_{I}$, $U_{II}$ realizes similar transformation
$\vert\psi_f\rangle = U_{II}\vert\psi_i\rangle$ on the subspace spanned by
$\vert\uparrow\downarrow\rangle$, $\vert\downarrow\uparrow\rangle$ and acts as a unit operator on the subspace spanned by $\vert\uparrow\uparrow\rangle$,
$\vert\downarrow\downarrow\rangle$. Let us consider quantum brachistochrone problem on each subspace in detail.

\subsection{Quantum evolution on the subspace spanned by $\vert\uparrow\uparrow\rangle$, $\vert\downarrow\downarrow\rangle$ \label{subsec3_1}}

We consider the quantum brachistochrone problem of two spins-$\frac{1}{2}$ represented by Hamiltonian (\ref{form8}) on the
$\vert\uparrow\uparrow\rangle$, $\vert\downarrow\downarrow\rangle$ subspace. As we noticed earlier Hamiltonian $H$ (\ref{form8}) acts on any state
$\vert\psi\rangle =a\vert\uparrow\uparrow\rangle + b\vert\downarrow\downarrow\rangle$ (where the normalization condition is the following:
$\vert a\vert ^2+\vert b\vert ^2=1$) as:
\begin{eqnarray}
H\vert\psi\rangle=H_I\vert\psi\rangle.\nonumber
\end{eqnarray}
Therefore, we consider the quantum brachistochrone problem on this subspace using Hamiltonian $H_I$ from (\ref{form8_1}).
Let us introduce the following operators:
\begin{eqnarray}
&&\sigma_x^I=\left( \begin{array}{ccccc}
0 & 0 & 0 & 1 \\
0 & 0 & 0 & 0 \\
0 & 0 & 0 & 0 \\
1 & 0 & 0 & 0
\end{array}\right),\quad
\sigma_y^I=\left( \begin{array}{ccccc}
0 & 0 & 0 & -i \\
0 & 0 & 0 & 0 \\
0 & 0 & 0 & 0 \\
i & 0 & 0 & 0
\end{array}\right),\nonumber\\
&&\sigma_z^I=\left( \begin{array}{ccccc}
1 & 0 & 0 & 0 \\
0 & 0 & 0 & 0 \\
0 & 0 & 0 & 0 \\
0 & 0 & 0 & -1
\end{array}\right),\quad
I^I=\left( \begin{array}{ccccc}
1 & 0 & 0 & 0 \\
0 & 0 & 0 & 0 \\
0 & 0 & 0 & 0 \\
0 & 0 & 0 & 1
\end{array}\right).
\label{f10}
\end{eqnarray}
These operators satisfy properties of Pauli and unit matrices for $\vert\uparrow\uparrow\rangle$, $\vert\downarrow\downarrow\rangle$ states when
we denote their as follows: $\vert\uparrow\uparrow\rangle\equiv\vert\uparrow\rangle$, $\vert\downarrow\downarrow\rangle\equiv\vert\downarrow\rangle$.

With the help of introduced operators (\ref{f10}) the Hamiltonian $H_{I}$ from (\ref{form8_1}) can be written in the form:
\begin{eqnarray}
H_I= \mbox{\boldmath{$ \sigma$}}^I \cdot {\bf h}^I + J_{zz} I^I,
\label{HI}
\end{eqnarray}
where ${\bf h}^I=(J_{Re}^-, J_{Im}^+, h^+)$. This Hamiltonian is similar to the Hamiltonian of one spin-$\frac{1}{2}$ in the magnetic field. In the
case of one spin-$\frac{1}{2}$, vector ${\bf h}^I$ is vector of the magnetic field. Brachistochrone problem for the spin-$\frac{1}{2}$ in the magnetic field
was considered in the paper \cite{ref4}. We cannot use the result from this paper because we assume another finite energy condition (\ref{energycond})
comparring to paper \cite{ref4}. In \cite{ref4} author fixed the largest and the smalest eigenvalues of the Hamiltonian.

Now, using (\ref{HI}) we represent the evolution operator $U_I=e^{-iH_It}$ as follows:
\begin{eqnarray}
U_I=\left(I-I^I+\cos(\omega_I t)I^I-\frac{i}{\omega_I}\sin(\omega_I t)\mbox{\boldmath{$ \sigma$}}^I \cdot {\bf h}^I\right)A^I,
\label{UI}
\end{eqnarray}
where
\begin{eqnarray}
A^I=\left( \begin{array}{ccccc}
e^{-iJ_{zz}t} & 0 & 0 & 0 \\
0 & 1 & 0 & 0 \\
0 & 0 & 1 & 0 \\
0 & 0 & 0 & e^{-iJ_{zz}t}
\end{array}\right).\nonumber
\end{eqnarray}
Here we use that $\left( \mbox{\boldmath{$ \sigma$}}^I \cdot {\bf h}^I \right)^{2n}={\omega_I}^{2n}I^I$ and
$\left(\mbox{\boldmath{$ \sigma$}}^I \cdot {\bf h}^I\right)^{2n+1}={\omega_I}^{2n} \mbox{\boldmath{$ \sigma$}}^I \cdot {\bf h}^I$ where $n=1,2,3,..$.
In a matrix form (\ref{UI}) reads:
\begin{eqnarray}
U_I=\left( \begin{array}{ccccc}
(\cos\left(\omega_It\right)-\frac{i}{\omega_I}\sin\left(\omega_It\right)h^+)e^{-iJ_{zz}t} & 0 & 0 & -\frac{i}{\omega_I}\sin\left(\omega_It\right)(J_{Re}^--iJ_{Im}^+)e^{-iJ_{zz}t} \\
0 & 1 & 0 & 0 \\
0 & 0 & 1 & 0 \\
-\frac{i}{\omega_I}\sin\left(\omega_It\right)(J_{Re}^-+iJ_{Im}^+)e^{-iJ_{zz}t} & 0 & 0 & (\cos\omega_It+\frac{i}{\omega_I}\sin\left(\omega_It\right)h^+)e^{-iJ_{zz}t}
\end{array}\right).
\label{formUI}
\end{eqnarray}

Let us put the initial state as $\vert\psi_i\rangle =\vert\uparrow\uparrow\rangle$ and the final one as
$\vert\psi_f\rangle =a\vert\uparrow\uparrow\rangle + b\vert\downarrow\downarrow\rangle$. Then using the matrix representation for the
operator of evolution $U_I$ (\ref{formUI}), the relation $\vert\psi_f\rangle =U_I\vert\psi_i\rangle$ takes the form:
\begin{eqnarray}
\left( \begin{array}{ccccc}
a \\
0 \\
0 \\
b
\end{array}\right)=e^{-iJ_{zz}t}
\left( \begin{array}{ccccc}
\cos\left(\omega_It\right)-\frac{i}{\omega_I}\sin\left(\omega_It\right)h^+ \\
0 \\
0 \\
-\frac{i}{\omega_I}\sin\left(\omega_It\right)(J_{Re}^-+iJ_{Im}^+)
\end{array}\right).
\label{evrel1}
\end{eqnarray}
From the fourth component of (\ref{evrel1}) we obtain the necessary condition that the initial state reaches the final one:
\begin{eqnarray}
\frac{1}{\omega_I}\sin\left(\omega_It\right)\sqrt{{J_{Re}^-}^2+{J_{Im}^+}^2}=\vert b\vert.
\label{necescond}
\end{eqnarray}
From equation (\ref{necescond}) we obtain the time required to transform the initial state $\vert\uparrow\uparrow\rangle$ into the final one
$a\vert\uparrow\uparrow\rangle + b\vert\downarrow\downarrow\rangle$:
$t=\frac{1}{\omega_I}\arcsin\left(\sqrt{1+\frac{{h^+}^2}{{J_{Re}^-}^2+{J_{Im}^+}^2}}\vert b\vert\right)$. Now, using the finite energy condition
(\ref{energyfinite}) we optimize the time of evolution if we put the following conditions:
\begin{eqnarray}
\omega_{II}=J_{zz}=h^+=0.
\label{condition1}
\end{eqnarray}
The optimal time is thus:
\begin{eqnarray}
\tau=\frac{1}{\omega}\arcsin\vert b\vert.
\label{form1time}
\end{eqnarray}
If the final state is $\vert\psi_f\rangle=\vert\downarrow\downarrow\rangle$ then we obtain the following passage time $\tau=\frac{\pi}{2\omega}$ (The shortest
time of evolution between the two fixed orthogonal states is called the passage time \cite{ref7}).

Condition (\ref{necescond}) does not account the phase of $b$.
Let us find the condition that allows to reach the phase of components of the final state during the optimal time (\ref{form1time}).
For this we insert the optimal time of evolution (\ref{form1time}) and conditions (\ref{condition1}) in equation (\ref{evrel1}).
From the fourth component of (\ref{evrel1}) we obtain the following equation: -$\frac{i}{\omega}\!\vert b\vert\!\left(\!J_{Re}^-\!+\!iJ_{Im}^+\!\right)\!=\!b$,
which allows to find another
conditions for optimal evolution: $J_{Re}^-=-\omega \frac{\Im b}{\vert b\vert}$, $J_{Im}^+=\omega \frac{\Re b}{\vert b\vert}$, where $b=\Re b+ i\Im b$.
Hence, all the necessary conditions for the optimal evolution in the subspace spanned by $\vert\uparrow\uparrow\rangle$, $\vert\downarrow\downarrow\rangle$ read:
\begin{eqnarray}
J_{xx}=-J_{yy}=-\frac{\omega}{2} \frac{\Im b}{\vert b\vert},\quad
J_{xy}=J_{yx}=\frac{\omega}{2} \frac{\Re b}{\vert b\vert},\quad
J_{zz}=0,\quad
h_z^1=h_z^2=0.
\label{cond1}
\end{eqnarray}
Initial state can reach maximally
entangled states $\frac{1}{\sqrt{2}}\left(\vert\uparrow\uparrow\rangle + e^{i\phi}\vert\downarrow\downarrow\rangle\right)$ ($\phi\in\left[0,2\pi\right]$)
during the minimal time $\tau =\frac{\pi}{4\omega}$.

If we put optimal conditions (\ref{cond1}) on the Hamiltonian $H$ (\ref{form8}) we obtain Hamiltonian which provides optimal evolution
on the subspace spanned by $\vert\uparrow\uparrow\rangle$, $\vert\downarrow\downarrow\rangle$. In the matrix form it reads:
\begin{eqnarray}
H=\left( \begin{array}{ccccc}
0 & 0 & 0 & -i\omega\frac{b^*}{\vert b\vert} \\
0 & 0 & 0 & 0 \\
0 & 0 & 0 & 0 \\
i\omega\frac{b}{\vert b\vert} & 0 & 0 & 0
\end{array}\right).
\label{optHI}
\end{eqnarray}

The Hilbert space of the $\vert\uparrow\uparrow\rangle$, $\vert\downarrow\downarrow\rangle$ subspace is two-dimentional and
it is represented by the Bloch sphere. Any pure state can be identified as a point on this sphere. The shortest distance between $\vert\psi_i\rangle$ and
$\vert\psi_f\rangle$ is a large circle arc on the sphere (geodesic path). The optimal way of transporting $\vert\psi_i\rangle$ into $\vert\psi_f\rangle$
is therefore to rotate the sphere around the axis ortogonal to the large circle. The axis of rotation passes along two quantum states
$\frac{1}{\sqrt{2}}\left[\vert\uparrow\uparrow\rangle + i\frac{b}{\vert b\vert}\vert\downarrow\downarrow\rangle\right]$,
$\frac{1}{\sqrt{2}}\left[\vert\uparrow\uparrow\rangle - i\frac{b}{\vert b\vert}\vert\downarrow\downarrow\rangle\right]$ which are eigenvectors
of optimal Hamiltonian $H$ (\ref{optHI}) that correspond to the largest $\omega$ and the smallest $-\omega$ eigenvalues.

\subsection{Quantum evolution on the subspace spanned by $\vert\uparrow\downarrow\rangle$, $\vert\downarrow\uparrow\rangle$ \label{subsec3_2}}

In this section we consider the quantum brachistochrone problem of two spin-$\frac{1}{2}$ represented by Hamiltonian (\ref{form8}) on the subspace
spaned by $\vert\uparrow\downarrow\rangle$, $\vert\downarrow\uparrow\rangle$. The Hamiltonian (\ref{form8}) acts on any state
$\vert\!\psi\!\rangle\! =\!a\vert\!\uparrow\downarrow\!\rangle + b\vert\downarrow\uparrow\rangle$ as:
\begin{eqnarray}
H\vert\psi\rangle=H_{II}\vert\psi\rangle.\nonumber
\end{eqnarray}
Let us introduse the following analogues Pauli and unit matrices for the basis states
$\vert\uparrow\downarrow\rangle\equiv\vert\uparrow\rangle$, $\vert\downarrow\uparrow\rangle\equiv\vert\downarrow\rangle$:
\begin{eqnarray}
&&\sigma_x^{II}=\left( \begin{array}{ccccc}
0 & 0 & 0 & 0 \\
0 & 0 & 1 & 0 \\
0 & 1 & 0 & 0 \\
0 & 0 & 0 & 0
\end{array}\right),\quad
\sigma_y^{II}=\left( \begin{array}{ccccc}
0 & 0 & 0 & 0 \\
0 & 0 & -i & 0 \\
0 & i & 0 & 0 \\
0 & 0 & 0 & 0
\end{array}\right),\nonumber\\
&&\sigma_z^{II}=\left( \begin{array}{ccccc}
0 & 0 & 0 & 0 \\
0 & 1 & 0 & 0 \\
0 & 0 & -1 & 0 \\
0 & 0 & 0 & 0
\end{array}\right),\quad
I^{II}=\left( \begin{array}{ccccc}
0 & 0 & 0 & 0 \\
0 & 1 & 0 & 0 \\
0 & 0 & 1 & 0 \\
0 & 0 & 0 & 0
\end{array}\right).
\label{f12}
\end{eqnarray}
Using (\ref{f12}) and making the same steps as in the previous case we rewrite Hamiltonian $H_{II}$ from (\ref{form8_1}) as:
\begin{eqnarray}
H_{II}= \mbox{\boldmath{$ \sigma$}}^{II} \cdot {\bf h}^{II} - J_{zz} I^{II},
\label{HII}
\end{eqnarray}
the components of vector ${\bf h}^{II}$ read ${\bf h}^{II}=(J_{Re}^+, -J_{Im}^-, h^-)$. The operator of evolution for this case $U_{II}=e^{-iH_{II}t}$
we can represent as:
\begin{eqnarray}
U_{II}=\left(I-I^{II}+\cos(\omega_{II} t)I^{II}-\frac{i}{\omega_{II}}\sin(\omega_{II} t)\mbox{\boldmath{$ \sigma$}}^{II} \cdot {\bf h^{II}}\right)A^{II},
\label{UII}
\end{eqnarray}
where
\begin{eqnarray}
A^{II}=\left( \begin{array}{ccccc}
1 & 0 & 0 & 0 \\
0 & e^{iJ_{zz}t} & 0 & 0 \\
0 & 0 & e^{iJ_{zz}t} & 0 \\
0 & 0 & 0 & 1
\end{array}\right).\nonumber
\end{eqnarray}
Here we use that $\left( \mbox{\boldmath{$ \sigma$}}^{II} \cdot {\bf h}^{II} \right)^{2n}={\omega_{II}}^{2n}I^{II}$ and
$\left(\mbox{\boldmath{$ \sigma$}}^{II} \cdot {\bf h}^{II}\right)^{2n+1}={\omega_{II}}^{2n} \mbox{\boldmath{$ \sigma$}}^{II} \cdot {\bf h}^{II}$
where $n=1,2,3,..$. In the matrix representation $U_{II}$ reads as:
\begin{eqnarray}
U_{II}=\left( \begin{array}{ccccc}
1 & 0 & 0 & 0 \\
0 & (\cos\left(\omega_{II}t\right)-\frac{i}{\omega_{II}}\sin\left(\omega_{II}t\right)h^-)e^{iJ_{zz}t} & -\frac{i}{\omega_{II}}\sin\left(\omega_{II}t\right)(J_{Re}^++iJ_{Im}^-)e^{iJ_{zz}t} & 0 \\
0 & -\frac{i}{\omega_{II}}\sin\left(\omega_{II}t\right)(J_{Re}^+-iJ_{Im}^-)e^{iJ_{zz}t} & (\cos\left(\omega_{II}t\right)+\frac{i}{\omega_{II}}\sin\left(\omega_{II}t\right)h^-)e^{iJ_{zz}t} & 0 \\
0 & 0 & 0 & 1
\end{array}\right).
\label{formUII}
\end{eqnarray}

Let us put the initial state as $\vert\psi_i\rangle =\vert\uparrow\downarrow\rangle$ and the final one as
$\vert\psi_f\rangle = a\vert\uparrow\downarrow\rangle + b\vert\downarrow\uparrow\rangle$. Similarly as in previous case using matrix
representation for $U_{II}$ (\ref{formUII}), we write the relation $\vert\psi_f\rangle =U_{II}\vert\psi_i\rangle$ in the form:
\begin{eqnarray}
\left( \begin{array}{ccccc}
0 \\
a \\
b \\
0
\end{array}\right)=e^{iJ_{zz}t}
\left( \begin{array}{ccccc}
0 \\
\cos\left(\omega_{II}t\right)-\frac{i}{\omega_{II}}\sin\left(\omega_{II}t\right)h^- \\
-\frac{i}{\omega_{II}}\sin\left(\omega_{II}t\right)(J_{Re}^+-iJ_{Im}^-) \\
0
\end{array}\right).
\label{evrel2}
\end{eqnarray}
In this case conditions for optimal evolution read:
\begin{eqnarray}
J_{xx}=J_{yy}=-\frac{\omega}{2} \frac{\Im b}{\vert b\vert},\quad
J_{xy}=-J_{yx}=-\frac{\omega}{2} \frac{\Re b}{\vert b\vert},\quad
J_{zz}=0,\quad
h_z^1=h_z^2=0,
\label{cond2}
\end{eqnarray}
and the optimal time is (\ref{form1time}).
Also, in this case, the initial state $\vert\uparrow\downarrow\rangle$ can reach maximally entangled states
$\frac{1}{\sqrt{2}}\left(\vert\uparrow\downarrow\rangle + e^{i\phi}\vert\downarrow\uparrow\rangle\right)$ ($\phi\in\left[0,2\pi\right]$)
during the minimal time $\tau =\frac{\pi}{4\omega}$.

If we put conditions (\ref{cond2}) on the Hamiltonian $H$ (\ref{form8}) we obtain Hamiltonian which provides optimal evolution
on the subspace spanned by $\vert\uparrow\downarrow\rangle$, $\vert\downarrow\uparrow\rangle$. In the matrix form this Hamiltonian reads:
\begin{eqnarray}
H=\left( \begin{array}{ccccc}
0 & 0 & 0 & 0  \\
0 & 0 & -i\omega\frac{b^*}{\vert b\vert} & 0 \\
0 & i\omega\frac{b}{\vert b\vert} & 0 & 0 \\
0 & 0 & 0 & 0
\end{array}\right).
\label{optHII}
\end{eqnarray}

The Hilbert space of the $\vert\uparrow\downarrow\rangle$, $\vert\downarrow\uparrow\rangle$ subspace is two-dimentional and
it is represented by the Bloch sphere. In this case the axis of rotation passes along two quantum states
$\frac{1}{\sqrt{2}}\left[\vert\uparrow\downarrow\rangle + i\frac{b}{\vert b\vert}\vert\downarrow\uparrow\rangle\right]$,
$\frac{1}{\sqrt{2}}\left[\vert\uparrow\downarrow\rangle - i\frac{b}{\vert b\vert}\vert\downarrow\uparrow\rangle\right]$ which are eigenvectors of the optimal
Hamiltonian $H$ (\ref{optHII}) that correspond the largest $\omega$ and the smallest $-\omega$ eigenvalues.

\section{Realization of quantum gates by two interacting spins \label{sec4}}

Operator of evolution can be related with some quantum gates. Target gate $U_{target}$ equals $U(T)$ modulo a global phase as:
\begin{eqnarray}
U(T)=e^{i\chi}U_{target},
\label{mglph}
\end{eqnarray}
where $T$ is the optimal time of generation of the target gate, $\chi$ is some real number.

We now demonstrate this explicitly by a few examples. Let us consider the entangler gate:
\begin{eqnarray}
U_{ENT}=\left( \begin{array}{ccccc}
\cos\phi & 0 & 0 & \sin\phi \\
0 & 1 & 0 & 0 \\
0 & 0 & 1 & 0 \\
-\sin\phi & 0 & 0 & \cos\phi
\end{array}\right),
\label{form14}
\end{eqnarray}
where $\phi \in [0,\pi]$. If this gate acts on the state
$\vert\uparrow\uparrow\rangle$, it will produce the $\phi$-dependent entangled state
$\cos\phi\vert\uparrow\uparrow\rangle - \sin\phi\vert\downarrow\downarrow\rangle$. The comparison of (\ref{form131}) with (\ref{form14}) using (\ref{formUI})
and (\ref{formUII})
leads to the following set of parameters: $J_{xx}=J_{yy}=J_{zz}=0$, $J_{xy}=J_{yx}=-\frac{\omega}{2}$,
$h_z^1=h_z^2=0$ and $t=\frac{\phi}{\omega}$. If we compare $t=\frac{\phi}{\omega}$ with optimal time (\ref{form1time}) we can see that this time is
the optimal time of evolution along the brachistochrone
on the subspace spanned by $\vert\uparrow\uparrow\rangle$, $\vert\downarrow\downarrow\rangle$ and the connection between parameter $\phi$
and $b$ is as follows $\phi=\arcsin \vert b\vert$ ($\phi$ is called the Wooters distance).

Entangled gate allow to reach Bell states from the nonentangled states. When $\phi=\frac{\pi}{4}$, this allows reaching the maximally entangled Bell state
$\vert\Phi^+\rangle =\frac{1}{\sqrt{2}}\left(\vert\uparrow\uparrow\rangle + \vert\downarrow\downarrow\rangle\right)$ from the initial state
$\vert\psi_I\rangle =\vert\downarrow\downarrow\rangle$ and Bell state
$\vert\Phi^-\rangle =\frac{1}{\sqrt{2}}\left(\vert\uparrow\uparrow\rangle - \vert\downarrow\downarrow\rangle\right)$
from the initial state $\vert\uparrow\uparrow\rangle$. Time of evolution between these pairs of states is the shortest possible time $t=\frac{\pi}{4\omega}$
which is allowed by Hamiltonian (\ref{form8}) and finite energy condition (\ref{energyfinite}).
Initial state $\vert\uparrow\uparrow\rangle$ can reach ortogonal final states $\vert\downarrow\downarrow\rangle$ during the
time $t=\frac{\pi}{2\omega}$, which is the passage time.
In \cite{ref51} the entangler gate is generated from operator of evolution which is obtained from Heisenberg Hamiltonian with $J_{ii}$ interaction between
spins ($i=x,y,z$) and magnetic field $h_z^{\alpha}$ ($\alpha =1,2$ for the first and second spin, respectively) which is directed along the $z$-axis.
In addition to this Hamiltonian, our Hamiltonian (\ref{form8}) contains items with $J_{xy}$ and $J_{yx}$ interaction which allow to generate the entangler
gate along a geodesic in the subspace spaned by $\vert\uparrow\uparrow\rangle$, $\vert\downarrow\downarrow\rangle$.

Let us construct a similar gate which produces the entangled states on the subspace spanned by $\vert\uparrow\downarrow\rangle$,
$\vert\downarrow\uparrow\rangle$. This gate is constructed with the entangler gate ($U_{ENT}$) and NOT gate
($U_{NOT}=\left( \begin{array}{ccccc}
0 & 1 \\
1 & 0 \\
\end{array}\right)$) as:
\begin{eqnarray}
U_{\overline{ENT}} = U_{NOT}^1 U_{ENT} U_{NOT}^1=\left( \begin{array}{ccccc}
1 & 0 & 0 & 0 \\
0 & \cos\phi & -\sin\phi & 0 \\
0 & \sin\phi & \cos\phi & 0 \\
0 & 0 & 0 & 1
\end{array}\right),
\label{form15}
\end{eqnarray}
where $U_{NOT}^1=U_{NOT}\otimes I$.
If gate (\ref{form15}) acts on the initial state $\vert\uparrow\downarrow\rangle$ we get entangled state
$\cos\phi\vert\uparrow\downarrow\rangle + \sin\phi\vert\downarrow\uparrow\rangle$. Similarly to the previous example we compare
unitary operator (\ref{form131}) with quantum gate (\ref{form15}) and obtain the following set of parameters:
$J_{xx}=J_{yy}=J_{zz}=0$, $J_{xy}=-J_{yx}=-\frac{\omega}{2}$, $h_z^1=h_z^2=0$ and $t =\frac{\phi}{\omega}$.
Time of generation of $U_{\overline{ENT}}$ is the optimal time of evolution along the brachistochrone (\ref{form1time})
($\phi=\arcsin \vert b\vert$) in the subspace spanned by $\vert\uparrow\downarrow\rangle$, $\vert\downarrow\uparrow\rangle$.

Let us insert $\phi=\frac{\pi}{4}$ in $U_{\overline{ENT}}$ (\ref{form15}) and act by $U_{\overline{ENT}}$ on the initial nonentangled state
$\vert\uparrow\downarrow\rangle$ or $\vert\downarrow\uparrow\rangle$.
We reach the maximally entangled Bell state
$\vert\Psi^+\rangle =\frac{1}{\sqrt{2}}\left(\vert\uparrow\downarrow\rangle + \vert\downarrow\uparrow\rangle\right)$ or the state that discribes EPR pair
$\vert\Psi^-\rangle =\frac{1}{\sqrt{2}}\left(\vert\uparrow\downarrow\rangle - \vert\downarrow\uparrow\rangle\right)$, respectively. In this case,
the time of evolution between these states is $t=\frac{\pi}{4\omega}$.
Evolution between two ortogonal states $\vert\uparrow\downarrow\rangle$ and $\vert\downarrow\uparrow\rangle$ is realized within the passage time
$t=\frac{\pi}{2\omega}$.

The next important gate which we consider is the SWAP gate:
\begin{eqnarray}
U_{SWAP}=\left( \begin{array}{ccccc}
1 & 0 & 0 & 0 \\
0 & 0 & 1 & 0 \\
0 & 1 & 0 & 0 \\
0 & 0 & 0 & 1
\end{array}\right),
\label{gate1}
\end{eqnarray}
which exchanges the states of two qubits. Similar steps as we did in case for entangler gate, we obtain the following set of parameters:
$J_{xx}=J_{yy}=J_{zz}=(-1)^p\frac{\omega}{\sqrt{6}}$, $J_{xy}=J_{yx}=0$, $h_z^1=h_z^2=0$, $t =\frac{\sqrt{6}\pi}{4\omega}$ and
$\chi =-\frac{\pi}{4}(-1)^p$ (where $p=0,1.$).
These parameters are the optimal conditions for $H$ (\ref{form8}) to generate SWAP gate. In other words, they allow to povide the unitary transformation
between $\vert\uparrow\downarrow\rangle$, $\vert\downarrow\uparrow\rangle$ states during the time $t =\frac{\sqrt{6}\pi}{4\omega}$. Similarly as in \cite{ref51}
the time-optimal generation of $U_{SWAP}$ is along a geodesic on the subspace spaned by $\vert\uparrow\downarrow\rangle$, $\vert\downarrow\uparrow\rangle$.

As in the previous example we can generate the gate which exchanges the
$\vert\uparrow\uparrow\rangle$ state into the
$\vert\downarrow\downarrow\rangle$
and vice versa:
\begin{eqnarray}
U_{\overline{SWAP}}=U_{NOT}^1 U_{SWAP} U_{NOT}^1 =\left( \begin{array}{ccccc}
0 & 0 & 0 & 1 \\
0 & 1 & 0 & 0 \\
0 & 0 & 1 & 0 \\
1 & 0 & 0 & 0
\end{array}\right),
\label{gate2}
\end{eqnarray}
when we put the following conditions: $J_{xx}=-J_{yy}=-J_{zz}=(-1)^p\frac{\omega}{\sqrt{6}}$, $J_{xy}=J_{yx}=0$, $h_z^1=h_z^2=0$,
$t =\frac{\sqrt{6}\pi}{4\omega}$ on unitary operator (\ref{form131}). Here $\chi = -\frac{\pi}{4}(-1)^p$ (where $p=0,1.$).
In this case, these condition allow to provide the unitary transformation between $\vert\uparrow\uparrow\rangle$, $\vert\downarrow\downarrow\rangle$
states during $t =\frac{\sqrt{6}\pi}{4\omega}$. The time-optimal evolution is along a geodesic on the subspace which is spaned by
$\vert\uparrow\uparrow\rangle$, $\vert\downarrow\downarrow\rangle$.

As a last example, we consider the unitary operator $U_{iSWAP}$:
\begin{eqnarray}
U_{iSWAP}=\left( \begin{array}{ccccc}
1 & 0 & 0 & 0 \\
0 & 0 & i & 0 \\
0 & i & 0 & 0 \\
0 & 0 & 0 & 1
\end{array}\right).
\label{form16}
\end{eqnarray}
It can be obtained from the time unitary operator (\ref{form131}) if
we put the following conditions: $J_{xx}=J_{yy}=-\frac{\omega}{2}$, $J_{zz}=J_{xy}=J_{yx}=0$, $h_z^1=h_z^2=0$ and $t =\frac{\pi}{2\omega}$.

Also, we can obtain the following gate:
\begin{eqnarray}
U_{\overline{iSWAP}} = U_{NOT}^1 U_{iSWAP} U_{NOT}^1=\left( \begin{array}{ccccc}
0 & 0 & 0 & i \\
0 & 1 & 0 & 0 \\
0 & 0 & 1 & 0 \\
i & 0 & 0 & 0
\end{array}\right),
\label{form17}
\end{eqnarray}
which is constructed using $U_{iSWAP}$ and $U_{NOT}$. This gate can be obtained if we put the following conditions:
$J_{xx}=-J_{yy}=-\frac{\omega}{2}$, $J_{zz}=J_{xy}=J_{yx}=0$, $h_z^1=h_z^2=0$ and $t=\frac{\pi}{2\omega}$ on unitary operator (\ref{form131}).

The value of the global phase $\chi$ for $U_{ENT}$, $U_{\overline{ENT}}$, $U_{iSWAP}$ and $U_{\overline{iSWAP}}$ gates is equal $2\pi p$, where
$p$ is arbitrary integer.

\section{Summary and Discussion \label{sec5}}

We have studied the quantum brachistochrone problem for the system of two-qubits represented by two spins interacting via anisotropic couplings
$J_{ii}$ ($i=x,y,z$), $J_{jk}$ ($j\neq k=x,y$) and magnetic fields $h_z^{\alpha}$ ($\alpha =1, 2$ for the first and second spin, respectively)
which is directed along the $z$-axis (\ref{form8}). This Hamiltonian
realizes quantum evolution in two subspaces spanned by $\vert\uparrow\uparrow\rangle$, $\vert\downarrow\downarrow\rangle$ and
$\vert\uparrow\downarrow\rangle$, $\vert\downarrow\uparrow\rangle$ and does not mix these subspaces because it does not contain items with
$\sigma_1^x\sigma_2^z$, $\sigma_1^z\sigma_2^x$ and $\sigma_1^y\sigma_2^z$, $\sigma_1^z\sigma_2^y$.

We solved this problem with the finite energy condition (\ref{energyfinite}) for each subspace separately. We obtained the conditions for optimal
quantum evolution in each subspace and calculated the shortest possible time for evolution from the initial
state $\vert\psi_i\rangle$ to the final one $\vert\psi_f\rangle$. Also, we obtained Hamiltonians (\ref{optHI}) and (\ref{optHII}) which provide optimal
evolution on the subspaces spanned by $\vert\uparrow\uparrow\rangle$, $\vert\downarrow\downarrow\rangle$ and
$\vert\uparrow\downarrow\rangle$, $\vert\downarrow\uparrow\rangle$, respectively.

The Hilbert space of each subspaces is two dimentional and it represented by the Bloch sphere.
The optimal way of transporting $\vert\psi_i\rangle$ into $\vert\psi_f\rangle$ is therefore to rotate the sphere around the axis which pass along the
eigenvectors that correspond the largest $\omega$ and the smallest $-\omega$ eigenvalues of the optimal Hamiltonian. Thus, the shortest distance
between $\vert\psi_i\rangle$ and $\vert\psi_f\rangle$  is a large
circle arc on the sphere (geodesic path).

We used our result for important examples of $U_{ENT}$, $U_{\overline{ENT}}$, $U_{SWAP}$, $U_{\overline{SWAP}}$, $U_{iSWAP}$ and
$U_{\overline{iSWAP}}$ gates. The authors of \cite{ref51} studied the time-optimal evolution of a unitary operator in context of variational
principle. In the two-qubit demonstration of their method they obtain optimal solution for generation some target quantum gates, namely the swap of qubits,
the quantum Fourier transformation and the entangler gate. In \cite{ref51} the time-optimal generation of the entangler gate does not occur along a geodesic
on the subspace spaned by $\vert\uparrow\uparrow\rangle$, $\vert\downarrow\downarrow\rangle$. In addition to Hamiltonian, which is considered in \cite{ref51},
our Hamiltonian (\ref{form8}) contains items with $J_{xy}$ and $J_{yx}$ interaction which allow to generate the entangler
gate along a geodesic. Also, we considered gate $U_{\overline{ENT}}$ (\ref{form15}) which allows to produce the entangler states on the subspace
$\vert\uparrow\downarrow\rangle$, $\vert\downarrow\uparrow\rangle$. These gates allow to reach maximal entangled Bell states from the nonentangled ones
during the shortest time $t =\frac{\pi}{2\omega}$.

Similarly as in \cite{ref51} we obtained the optimal parameters for $H$ (\ref{form8}) which allow to generate SWAP gate. Also, we generated $\overline{SWAP}$
gate (\ref{gate2}) which exchanges the $\vert\uparrow\uparrow\rangle$ into $\vert\downarrow\downarrow\rangle$ and vice versa.

The Hamiltonian for generation iSWAP gate was proposed in \cite{ref17}. The
authors showed that iSWAP operation can be obtain by applying $XY$ Hamiltonian with $-\frac{E^{XY}}{4}$ interacting couplings for a time
$t=\frac{\pi}{E^{XY}}$. Recently, in \cite{ref23} using analitical and numerical calculations it were got minimal times required for various quantum gates
such as iSWAP, controled-$\pi$-phase (or CZ), CNOT, $\sqrt{SWAP}$ and Toffoli gates under typical experimental conditions. For creating iSWAP gate they
consider a physical system of two-qubit describing by Heisenberg Hamiltonian with $J_{zz}$ and $J_{yy}$ interacting couplings, which is equal with each other
$J_{zz}=J_{yy}=\frac{J}{2}$, in magnetic field $-\frac{\Delta}{2}$ directed along $x$-axis. This
Hamiltonian effects an iSWAP gate, in addition to two single-qubit rotations, during the time $t=\frac{\pi}{2J}$.
In addition to results of previous works we obtained the conditions for $H$ (\ref{form8}) to generate of
$\overline{iSWAP}$ gates (\ref{form17}). This gate does not change states from $\vert\uparrow\downarrow\rangle$, $\vert\downarrow\uparrow\rangle$ subspace
and realizes the following transformations: $U_{\overline{iSWAP}}\vert\uparrow\uparrow\rangle = i\vert\downarrow\downarrow\rangle$,
$U_{\overline{iSWAP}}\vert\downarrow\downarrow\rangle = i\vert\uparrow\uparrow\rangle$ with states from another subspase.

\section{Acknowledgment}

Authors would like to thank Dr. A. Rovenchak and Dr. M. Stetsko for useful comments.


\begin{thebibliography}{99}
\bibitem{ref1} Alberto Carlini, Akio Hosoya, Tatsuhiko Koike and Yosuke Okudaira, Phys. Rev. Lett. \textbf{96}, 060503 (2006).
\bibitem{ref6} Dorje C. Brody, and Daniel W. Hook, J. Phys. A \textbf{39}, L167 (2006).
\bibitem{ref2} Carl M. Bender, Dorje C. Brody, Hugh F. Jones and Bernhard K. Meister, Phys. Rev. Lett. \textbf{98}, 040403 (2007).
\bibitem{ref3} Carl M. Bender, and Dorje C. Brody, Time in Quantum Mechanics --
\textbf{2}, Lecture Notes in Physics, \textbf{789}, Springer: Berlin -- Heidelberg, 341 (2010).
\bibitem{ref31} Carl M. Bender, SIGMA \textbf{3}, 126 (2007).
\bibitem{ref4} V. M. Tkachuk, Fundamental problems of quantum mechanic (Ivan Franko National University of Lviv, Lviv (2011)). [in Ukrainian]
\bibitem{ref7} Dorje C. Brody, J. Phys. A \textbf{36}, 5587 (2003).
\bibitem{ref52} Alberto Carlini, Akio Hosoya, Tatsuhiko Koike and Yosuke Okudaira, arXiv:quant-ph/0703047.
\bibitem{ref8} A. M. Frydryszak, V. M. Tkachuk, Phys. Rev. A \textbf{77}, 014103 (2008).
\bibitem{ref9} V. Giovannetti, S. Lloyd and L. Maccone, Europhys. Lett. \textbf{62}, 615 (2003).
\bibitem{ref10} V. Giovannetti, S. Lloyd and L. Maccone, Phys. Rev. A \textbf{67}, 052109 (2003).
\bibitem{ref11} J. Batle, M. Casas, A. Plastino and A. R. Plastino,
Phys. Rev. A \textbf{72}, 032337 (2005).
\bibitem{ref12} A. Borras, M. Casas, A. R. Plastino and A. Plastino,
Phys. Rev. A \textbf{74}, 022326 (2006).
\bibitem{ref13} C. Zander, A. R. Plastino, A. Plastino and M. Casas,
J. Phys. A \textbf{40}, 2861 (2007).
\bibitem{ref14} A. Borras, C. Zander, A. R. Plastino, M. Casas and A. Plastino, Europhys. Lett.
\textbf{81}, 30007 (2008).
\bibitem{ref15} A. Borras, A. R. Plastino, M. Casas and A. Plastino, Phys. Rev. A \textbf{78}, 052104 (2008).
\bibitem{ref16} Bao-Kui Zhao, Fu-Guo Deng, Feng-Shou Zhang and Hong-Yu Zhou, Phys. Rev. A \textbf{80}, 052106 (2009).
\bibitem{ref51} Alberto Carlini, Akio Hosoya, Tatsuhiko Koike and Yosuke Okudaira, Phys. Rev. A \textbf{75}, 042308 (2007).
\bibitem{ref23} S. Ashhab, P. C. de Croot and Franco Nori, Phys. Rev. A \textbf{85}, 052327 (2012).
\bibitem{ref231} N. Khaneja, R. Brockett and S. J. Glaser, Phys. Rev. A \textbf{63}, 032308 (2001).
\bibitem{ref232} G. Dirr, U. Helme, K. Hupr, M. Kleinsteuber and Y. Liu, J. Glob. Opt. \textbf{35}, 443 (2006).
\bibitem{ref233} G. Vidal, K. Hammerer and J. I. Cirac, Phys. Rev. Lett \textbf{88}, 237902 (2002).
\bibitem{ref234} G. Vidal, K. Hammerer and J. I. Cirac, Phys. Rev. A \textbf{66}, 062321 (2002).
\bibitem{ref235} L. Masanes, G. Vidal and J. I. Latorre, Quantum Inf. Comput. \textbf{2}, 285 (2002).
\bibitem{ref236} J. Zhang, J. Vala, S. Sastry and K. B. Whaley, Phys. Rev. A \textbf{67}, 042313 (2003).
\bibitem{ref24} N. Khaneja, S. J. Glaser and R. Brockett, Phys. Rev. A \textbf{65}, 032301 (2002).
\bibitem{ref25} T. O. Reiss, N. Khaneja, S. J. and Glaser, J. Magn. Reson \textbf{165}, 95 (2003).
\bibitem{ref26} H. Yuan and N. Khaneja, Phys. Rev. A \textbf{72}, 040301 (2005).
\bibitem{ref27} R. Zeier, H. Yuan and N. Khaneja, Phys. Rev. A \textbf{77}, 032332 (2008).
\bibitem{ref28} A. M. Childs, H. L. Haselgrove and M. A. Nielsen, Phys. Rev. A \textbf{68}, 052311 (2003).
\bibitem{ref29} R. Zeier, M. Grassl and T. Beth, Phys. Rev. A \textbf{70}, 032319 (2004).
\bibitem{ref30} T. Schulte-Herbuggen, A. Sporl, N. Khaneja and S. J. Glaser, Phys. Rev. A \textbf{72}, 042331 (2004).
\bibitem{ref311} M. A. Nielsen, M. Dowling, M. Gu and A. Doherty, Science \textbf{311}, 1133 (2006).
\bibitem{ref32} M. A. Nielsen, M. Dowling, M. Gu and A. Doherty, Phys. Rev. A \textbf{73}, 062323 (2006).
\bibitem{ref17} N. Schuch and J. Siewert, Phys. Rev. A \textbf{67}, 032301 (2003).
\bibitem{ref18} Adriano Barenco, Charles H. Bennett, Richard Cleve, David P. DiVincenzo, Norman Margolus, Peter Shor, Tycho Sleator, John A. Smolin and
Harald Weinfurter, Phys. Rev. A, \textbf{52}, 3457 (1995).
\bibitem{ref171} David P. DiVincenzo, Royal Society of London Proceeding Series A, \textbf{454}, 261 (1998).
\bibitem{ref22} Alberto Carlini, Akio Hosoya, Tatsuhiko Koike and Yosuke Okudaira, J. Phys. A, \textbf{44}, 145302 (2011).
\bibitem{ref33} J. Anandan and Y. Aharonov, Phys. Rev. Lett. \textbf{65}, 1697 (1990).
\end{thebibliography}
\end{document}